\documentclass[reprint,superscriptaddress,amsmath,amssymb,aps, prl, longbibliography]{revtex4-2}
\usepackage{dcolumn, comment, amsmath, amsfonts, physics, mathcomp, bm, mathtools}
\usepackage{color, float}
\usepackage{ascmac, fancybox, ulem}
\usepackage{placeins, tcolorbox}
\usepackage[dvipsnames]{xcolor} %comment out
\usepackage[colorlinks,citecolor=Violet,linkcolor=BrickRed]{hyperref}  %PineGreen,dvipdfmx
\begin{document}
\preprint{APS/123-QED}

\title{Kitaev Meets Affleck-Kennedy-Lieb-Tasaki: Competing Quantum Disorder in Spin-$3/2$ Honeycomb Systems}

\author{Sogen Ikegami}
\email{ikegami-sogen443@g.ecc.u-tokyo.ac.jp}
  \affiliation{Department of Applied Physics, The University of Tokyo, Tokyo 113-8656, Japan}
\author{Kiyu Fukui}%
  \affiliation{Department of Physical Sciences, Ritsumeikan University, Kusatsu, Shiga 525-8577, Japan}
\author{Rico Pohle}%
  \affiliation{Faculty of Science, Shizuoka University, Shizuoka 422-8529, Japan}
\author{Yukitoshi Motome}%
  \email{motome@ap.t.u-tokyo.ac.jp}
  \affiliation{Department of Applied Physics, The University of Tokyo, Tokyo 113-8656, Japan}

\date{\today}% It is always \today, today,

\begin{abstract}
We investigate an \mbox{$S=3/2$} quantum spin model on a two-dimensional honeycomb lattice 
that continuously interpolates between two paradigmatic quantum disordered states with distinct entanglement structures: 
the Kitaev quantum spin liquid and the Affleck-Kennedy-Lieb-Tasaki (AKLT) valence bond solid. 
Combining classical, semi-classical, and exact diagonalization approaches, 
we map out the ground-state phase diagram and
elucidate the role of quantum fluctuations across the entire parameter range.
While classical and semi-classical frameworks predict
noncoplanar orders competing with a collinear N\'eel state,
we find these phases to be fragile: once full quantum fluctuations are included, 
they melt into a quantum-entangled state characterized by
suppressed spin correlations and enhanced entanglement entropy.
Our findings highlight how competition between qualitatively different quantum disordered phases
provides a fertile playground for unconventional phases emerging from their interplay and quantum fluctuations.
\end{abstract}

                              %3750 words from abstract to acknowledgment

                              % Fig.1 : 1.88 aspect ratio, double-column -> 368words
                              % Fig.2 : 1.28 aspect ratio, single-column -> 139words
                              % Fig.3 : 1.61 aspect ratio, single-column -> 113words
                              % 620 words in total.
                              % equations: 16*3=48words

\maketitle

%\tableofcontents

%---------------------------- Introduction ----------------------------%
\textit{Introduction.---}
Quantum fluctuations often suppress long-range order even at zero temperature,
giving rise to various quantum disordered states.
Among the most prominent are quantum spin liquids (QSLs), which exhibit
long-range entanglement and fractionalized excitations \cite{Balents2010, Savary_2017, RevModPhys.89.025003}.
Their exotic topological properties, lying beyond the Landau-Ginzburg-Wilson paradigm,
have attracted considerable interest not only in frustrated magnetism
but also in quantum information science and quantum computation \cite{doi:10.1142/S0217979290000139, PhysRevB.41.9377, 10.1063/1.1499754, KITAEV20032, PhysRevLett.90.227902, KITAEV20062, PhysRevLett.96.110404, PhysRevLett.96.110405, RevModPhys.80.1083}.
Understanding the fundamental characteristics of quantum disordered phases
and exploring novel quantum phases remain central challenges in modern condensed matter physics.

Since P. W. Anderson's pioneering proposal of the resonating valence bond (RVB) state~\cite{ANDERSON1973153}, 
various theoretical frameworks have been developed to describe QSLs and their intrinsic properties 
\cite{PhysRevLett.59.2095, PhysRevLett.61.2376, PhysRevB.40.7387, PhysRevB.44.2664, KITAEV20032, KITAEV20062, PhysRevLett.113.197205, PhysRevB.109.134412}. 
Frustration arising from competing interactions plays a crucial role 
in stabilizing these quantum disordered states, and they are often
studied in regimes where different conventional ordered phases compete. 
While the competing ordered phases and the associated order-disorder transitions have been extensively studied,
much less is known about systems in which multiple,
qualitatively different quantum disordered states compete with each other. 
When such states with distinct entanglement structures coexist, 
their interplay can enhance quantum fluctuations and potentially stabilize
unconventional quantum phases.

In this study, we focus on a promising platform 
where competition between distinct quantum disordered states 
naturally arises---an \mbox{$S=3/2$} spin system on a two-dimensional honeycomb lattice. 
This system hosts two well-established quantum disordered states: 
the Kitaev QSL~\cite{KITAEV20062} and the Affleck-Kennedy-Lieb-Tasaki (AKLT) valence bond solid (VBS)~\cite{PhysRevLett.59.799, Affleck1988, Kennedy1988}.
Although the Kitaev QSL was originally proposed for \mbox{$S=1/2$} spins,
its essential features 
persist for higher spins; notably, the \mbox{$S=3/2$} Kitaev honeycomb model supports a gapless QSL ground state 
with fractionalized excitations and long-range entanglement \cite{PhysRevLett.98.247201, PhysRevB.78.115116, PhysRevLett.130.156701, PhysRevB.80.064413, PhysRevLett.102.217202, Jin2022, PhysRevB.108.075111}.
Similarly, the AKLT VBS state, first introduced for a one-dimensional \mbox{$S=1$} system, 
generalizes to higher-spin systems on various lattice geometries \cite{PhysRevLett.59.799, Affleck1988, Kennedy1988}. 
On the honeycomb lattice, the \mbox{$S=3/2$} AKLT state is constructed from three \mbox{$S=1/2$}
constituents forming nearest-neighbor singlets,
and is classified as a (weak) symmetry-protected topological phase with short-range entanglement \cite{PhysRevB.88.205124}.
Despite their fundamentally different entanglement structures,
both states are of significant interest for quantum information applications~\cite{KITAEV20032, KITAEV20062, RevModPhys.80.1083, PhysRevLett.106.070501, MIYAKE20111656, PhysRevA.90.042333}.

%374 words up to here.

\begin{figure*}[t]
    \hypertarget{fig:schematic_pd}{}
    \includegraphics[width=2\columnwidth]{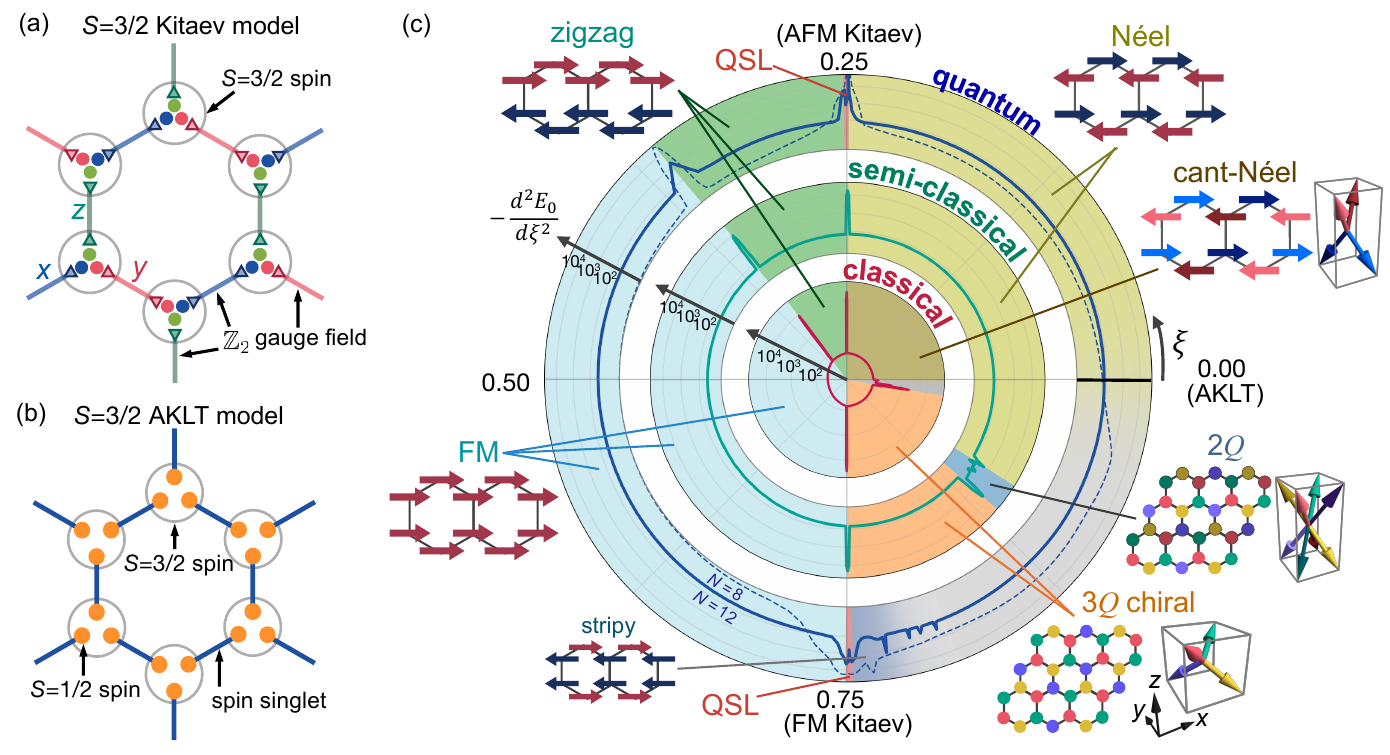}
    \caption{\label{fig:schematic_pd} 
    (a) Schematic illustration of the \mbox{$S=3/2$} Kitaev QSL on the honeycomb lattice in the \mbox{SO(6)} Majorana representation.
    Each \mbox{$S=3/2$} spin (large gray circle) is decomposed into three itinerant Majorana fermions (small circles) and three gauge Majorana fermions (triangles). 
    The gauge Majorana fermions define $\mathbb{Z}_2$ conserved quantities on each bond, indicated by red, green, and blue bonds.
    (b) Graphical representation of the AKLT VBS state on the honeycomb lattice.
    Each \mbox{$S=3/2$} spin (large gray circle) is decomposed into three \mbox{$S=1/2$} spins (small orange circles),
    and each pair of neighboring \mbox{$S=1/2$} spins forms a singlet state (blue line).
    Projecting the three \mbox{$S=1/2$} spins at each site onto the fully symmetric subspace yields
    the AKLT VBS state. 
    (c) Ground-state phase diagrams of the Kitaev-AKLT honeycomb model defined in Eq.~\eqref{eq:H} obtained by three complementary methods:
    classical \mbox{O(3)} vector analysis (inner circle), semi-classical \mbox{SU(4)} coherent state approach (middle circle), and full quantum ED (outer circle).
    Each circle displays the second derivative of the ground state energy \mbox{$E_0$} with respect to $\xi$, $-\frac{\partial^2 E_0}{\partial \xi^2}$.
    For ED, the solid and dashed lines correspond to the results for \mbox{$N=12$} and \mbox{$N=8$} clusters [see also the inset of Fig.~\protect\hyperlink{fig:EE}{\ref{fig:EE}(b)}], respectively;
    the \mbox{$N=8$} data are multiplied by a factor of 10 for visibility.
    See the text for details.
    }
    %229 words in caption
\end{figure*}

In this Letter, we investigate how these two quantum disordered phases compete 
by introducing a parameter that continuously interpolates between the Kitaev and AKLT limits. 
Using three complementary approaches,  
classical, semi-classical, and fully quantum, 
we systematically elucidate the ground-state phase diagram and clarify the impact of quantum fluctuations.
At classical and semi-classical levels, 
in addition to ferromagnetic (FM), antiferromagnetic (AFM) N\'eel, 
and zigzag ordered phases in weakly competing regions,
we identify noncoplanar orders in the regime where the
Kitaev QSL and the AKLT VBS compete.
However, these phases are progressively 
destabilized as quantum fluctuations are incorporated;
in the fully quantum results, this regime instead
exhibits suppressed spin correlations
and enhanced entanglement entropy---hallmarks of a quantum-entangled disordered state.
Our findings provide new insights into the interplay between quantum disordered states,
offering fresh perspectives on the organizing principles of exotic quantum phases.

%150 words in this paragraph

%---------------------------- Model ----------------------------%
\textit{Model Hamiltonian.---}
  We consider an \mbox{$S=3/2$} quantum spin model on the honeycomb lattice, whose Hamiltonian 
  interpolates between the Kitaev and AKLT limits:
  \begin{equation}\label{eq:H}
    \hat{H} = \sin(2\pi \xi)\hat{H}_{\rm{Kitaev}} + \cos(2\pi \xi) \hat{H}_{\rm{AKLT}},
  \end{equation}
  where \mbox{$\hat{H}_{\rm{Kitaev}}$} and \mbox{$\hat{H}_{\rm{AKLT}}$} denote
  the \mbox{$S=3/2$} Kitaev and AKLT Hamiltonians, respectively,
  and the parameter \mbox{$\xi \in [0,1]$} interpolates between the two limits.
  The Kitaev term in Eq.~\eqref{eq:H} is explicitly given by
  \begin{equation}\label{eq:HKitaev}
    \hat{H}_{\rm{Kitaev}} = \sum_{\gamma= x,y,z} \sum_{\langle i,j \rangle_\gamma}K\hat{S}_i^\gamma \hat{S}_j^\gamma,
  \end{equation}
  where \mbox{$\hat{S}_i^\gamma$} ($\gamma = x,y,z$) represents the $\gamma$ component of the \mbox{$S=3/2$} spin operator at site $i$, 
  and \mbox{$\langle i,j \rangle_\gamma$} denotes nearest-neighbor pairs on the $\gamma$-type bonds of the honeycomb lattice,
  as illustrated by blue, red, and green for $x$, $y$, and $z$ 
  in Fig.~\hyperlink{fig:schematic_pd}{\ref{fig:schematic_pd}(a)}.
  An \mbox{SO(6)} Majorana representation of \mbox{$S=3/2$} spins 
  introduces static $\mathbb{Z}_2$ gauge fields and three flavors of itinerant fermions,
  resulting in a gapless QSL \cite{PhysRevB.80.064413, PhysRevLett.102.217202, Jin2022, PhysRevB.108.075111}.
  The AKLT term in Eq.~\eqref{eq:H} is defined as
  \begin{equation}\label{eq:HAKLT}
    \hat{H}_{\rm{AKLT}} = \sum_{\langle i,j \rangle}J\left[ \hat{\bm{S}}_i \cdot \hat{\bm{S}}_j + \frac{116}{243}(\hat{\bm{S}}_i \cdot \hat{\bm{S}}_j)^2 + \frac{16}{243}(\hat{\bm{S}}_i \cdot \hat{\bm{S}}_j)^3 \right],
  \end{equation}
  where \mbox{$\hat{\bm{S}}_i=(\hat{S}^x_i, \hat{S}^y_i, \hat{S}^z_i)$}.
  This is equivalently reformulated in terms of the projection operator onto the total-spin
  \mbox{$S=3$} subspace for each nearest-neighbor pair as
  \mbox{$\hat{H}_{\rm{AKLT}} =\sum_{\langle i,j \rangle}J\left(\frac{160}{27}\hat{P}_{ij}^{S=3} - \frac{55}{108}\right)$},
  where the projection operator \mbox{$\hat{P}_{ij}^{S=3}$} takes the eigenvalue of $+1$ when the total spin of
  neighboring spins at site $i$ and $j$ is $S=3$, and $0$ otherwise.
  At \mbox{$\xi=0$} and \mbox{$1$} in Eq.~\eqref{eq:H}, the model reduces to the AKLT Hamiltonian,
  whose exact ground state is the AKLT VBS.
  In contrast, at \mbox{$\xi=0.25$} and \mbox{$0.75$}, the model corresponds to the AFM and FM \mbox{$S=3/2$} Kitaev models, respectively,
  both realizing a QSL ground states.
  We set \mbox{$K=J=1$}.
  We note that this Kitaev-AKLT model was recently studied on an \mbox{$S=1$} spin chain~\cite{raja2025kitaevakltmodel}.

  % 341 words

%---------------------------- Methods ----------------------------%
\textit{Three complementary approaches.---}
  To explore the ground state properties across the entire interpolation parameter \mbox{$\xi$}, 
  we employ three complementary methods that incorporate quantum fluctuations at different levels:
  (i) classical \mbox{O(3)} framework, (ii) semi-classical \mbox{SU(4)} coherent-state approach, and (iii) exact diagonalization (ED).
  In the classical \mbox{O(3)} framework, \mbox{$S=3/2$} spins are treated as classical vectors of fixed length \mbox{$|\bm{S}_i|=3/2$}.
  In contrast, the semi-classical approach represents each \mbox{$S=3/2$} spin by an \mbox{SU(4)} coherent state, 
  fully capturing onsite quantum fluctuations while neglecting intersite quantum entanglement
  \cite{Perelomov1972, Gnutzmann_1998, Nemoto_2000, READ1989609, PhysRevB.79.214436, PhysRevB.104.104409, PhysRevB.107.L140403, PhysRevB.108.L241108, pohle2025electronphonon}.
  For both calculations, we perform large-scale energy optimization using a gradient descent method
  \cite{kingma2017adam, jax2018github, deepmind2020jax}.
  Further details of these methods are provided in the Supplemental Material~\cite{SM}.
  Although these approaches cannot fully capture quantum-entangled states
  such as the Kitaev QSL or the AKLT VBS,
  they provide valuable insights into the global phase diagram and the role of quantum fluctuations. 
  To fully account for quantum fluctuations, 
  we perform ED calculations using the Lanczos algorithm on clusters of up to \mbox{$N=12$} sites under periodic boundary conditions.

  %184words

%---------------------------- Result: Ground-state phase diagrams ----------------------------%
\textit{Ground-state phase diagrams.---}
  Figure~\hyperlink{fig:schematic_pd}{\ref{fig:schematic_pd}(c)} summarizes
  the ground-state phase diagrams of the Kitaev-AKLT model in Eq.~\eqref{eq:H} obtained by the three approaches.
  The inner, middle, and outer circles correspond to the classical, semi-classical, and ED results, respectively.
  To identify phase boundaries, we compute the second derivative of the ground state energy 
  \mbox{$E_0$} with respect to \mbox{$\xi$},
  and locate its peaks as indicators of phase transitions.
  Each phase is identified by spin structures of the corresponding ground state.

  In the left half of the phase diagram \mbox{($0.25 < \xi < 0.75$)}, 
  FM and zigzag orders dominate consistently across all three approaches, 
  with narrow regions near \mbox{$\xi=0.25$} and \mbox{$0.75$} identified as the Kitaev QSLs in the ED results. 
  Conversely, in the right half \mbox{($0.00 < \xi < 0.25$} and \mbox{$0.75 < \xi < 1.00$)}, the results exhibit strong approach dependence.
  The classical framework shows a canted N\'eel order with slight noncoplanar canting %from the collinear N\'eel configuration 
  for \mbox{$0.00 < \xi <0.25$}, 
  and a noncoplanar \mbox{triple-$Q$} (3$Q$) order with nonzero spin scalar chirality for \mbox{$0.75 < \xi \lesssim 0.98$} \cite{PhysRevLett.79.2081, doi:10.1143/JPSJ.79.083711, PhysRevB.83.184401, PhysRevResearch.6.033077}; see Supplemental Material~\cite{SM}. 
  The semi-classical approach stabilizes a collinear N\'eel phase over \mbox{$0.90 \lesssim \xi < 1.00$} and \mbox{$0.00 < \xi <0.25$}, 
  while reducing the \mbox{$3Q$} chiral region to \mbox{$0.75 < \xi \lesssim 0.88$} and leaving a narrow window of a distinct \mbox{double-$Q$} \mbox{($2Q$)} phase with vanishing chirality.
  Stability of the N\'eel, FM, and zigzag orders observed at the semi-classical level is
  further supported by the generalized spin-wave theory (GSWT) \cite{10.1093/ptep/ptu109} (see Supplemental Material~\cite{SM}).
  In contrast, the full quantum ED calculations strongly suppress these noncoplanar orders, 
  leaving no clearly identifiable magnetic orders in \mbox{$0.80 \lesssim \xi < 1.00$}, 
  where the FM Kitaev QSL near \mbox{$\xi = 0.75$} and the AKLT VBS near \mbox{$\xi = 1.00$} compete~\cite{ED_note}.

  %330 words

  \begin{figure}[t]
  \hypertarget{fig:EE}{}
  \includegraphics[width=\columnwidth]{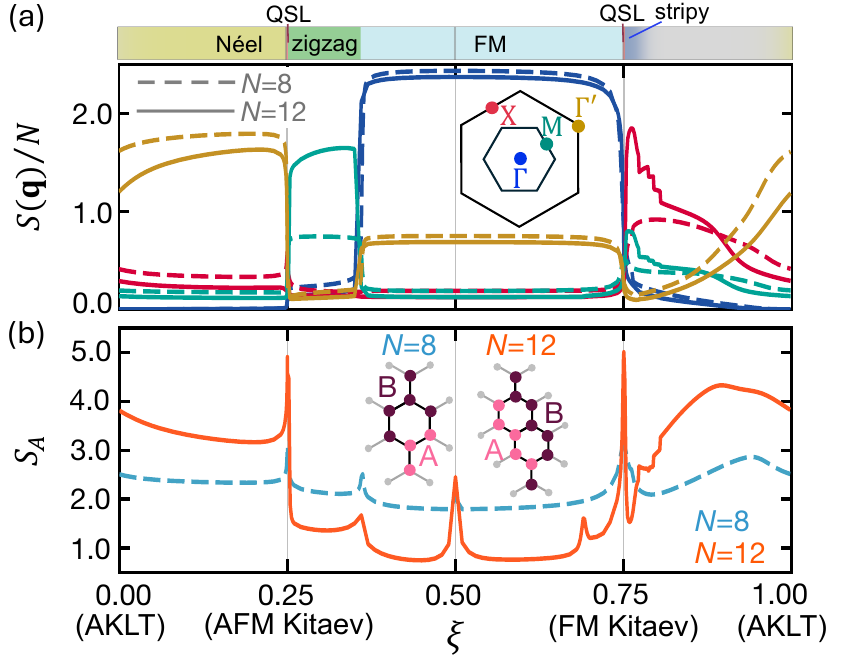}
  \caption{\label{fig:EE} 
  $\xi$ dependence of (a) the normalized spin structure factor $S(\mathbf{q})/N$ at the $\Gamma$, M, X, and $\Gamma^\prime$ points
  and (b) the entanglement entropy (EE),
  obtained from ED for \mbox{$N=8$} and \mbox{$N=12$} clusters.
  The inset of (a) represents the Brillouin zone of the honeycomb lattice,
  with the inner and outer hexagons indicating the first Brillouin zone 
  and the extended one up to the third zone, respectively. 
  The inset of (b) illustrates the cluster shapes and bipartition scheme used for the EE calculations. 
  %84
  }
  \end{figure}

  To further elucidate the quantum phases suggested by ED,
  Fig.~\hyperlink{fig:EE}{\ref{fig:EE}} shows the $\xi$ dependence of
  the normalized spin structure factor at several high-symmetry points in momentum space
  and the entanglement entropy (EE),
  obtained for \mbox{$N=8$} and \mbox{$N=12$} clusters.
  The spin structure factor is defined as
  \mbox{$S(\mathbf{q}) = \frac{1}{N} \sum_{i,j} \expval{\hat{\bm{S}}_i \cdot \hat{\bm{S}}_j}{\text{GS}} e^{i \mathbf{q} \cdot (\bm{r}_i - \bm{r}_j)}$},
  where $\bm{r}_i$ represents the position vector of site $i$.
  The EE is defined as \mbox{$S_A = -{\rm{Tr}}_A [\rho_A \log \rho_A]$} with 
  \mbox{$\rho_A = {\rm{Tr}}_B [\ket{\text{GS}} \bra{\text{GS}}]$},
  where $\ket{\text{GS}}$ represents the ground-state wavefunction, and \mbox{${\rm{Tr}}_{A(B)}$} denotes the partial trace over subsystem $A(B)$.
  The bipartition scheme is illustrated in the inset of Fig.~\hyperlink{fig:EE}{\ref{fig:EE}(b)}. %+10
  Small EE indicates that the classical and semi-classical approaches are good approximations,  
  while large EE signals significant quantum entanglement beyond these approximations.

  For \mbox{$0.25 \lesssim \xi \lesssim 0.75$},
  $S(\mathbf{q})/N$ exhibits strong intensities corresponding to the FM and zigzag orders, 
  and the relatively small EE supports that the classical and semi-classical results
  in Fig.~\hyperlink{fig:schematic_pd}{\ref{fig:schematic_pd}(c)} well capture the ground state of this model.
  The peak in EE at \mbox{$\xi=0.5$} is attributed to the higher SU(2) symmetry of the model, 
  while the origin of another one at \mbox{$\xi \sim 0.7$} is not fully understood at present. %+30words
  In contrast, the large EE for both clusters in \mbox{$0.00 \lesssim \xi \lesssim 0.25$} and \mbox{$0.75 \lesssim \xi \lesssim 1.00$}
  indicates quantum-entangled states beyond the semi-classical picture.
  In particular, pronounced peaks in EE near the AFM \mbox{($\xi=0.25$)} and FM \mbox{($\xi=0.75$)} Kitaev limits reflect the entangled nature of the Kitaev QSLs.
  While the intensity of $S(\mathbf{q}=\Gamma')/N$ suggests N\'eel correlations in \mbox{$0.00 \lesssim \xi \lesssim 0.25$}, 
  consistent with the classical and semi-classical predictions, it is progressively suppressed beyond the AKLT point. 
  In the region \mbox{$0.75\lesssim \xi \lesssim 1.00$}, where
  the AKLT VBS and the FM Kitaev QSL compete, 
  the intensities of $S(\mathbf{q})/N$ become generally weak---except 
  at the X points near \mbox{$\xi \sim 0.75$} (see below).
  In the same region, the EE forms a broad peak with values comparable to those of the Kitaev QSLs.

\begin{figure}[t]
  \hypertarget{fig:str-fac}{}
  \includegraphics[width=\columnwidth]{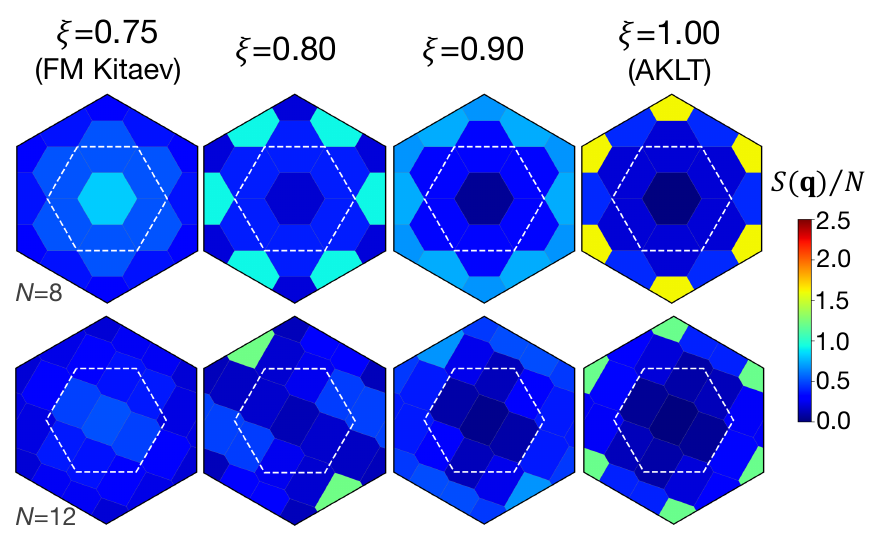}
  \caption{\label{fig:str-fac} 
  Momentum space distributions of $S(\mathbf{q})/N$ up to the third Brillouin zone for $\xi=0.75$, $0.80$, $0.90$, and $1.00$ 
  on \mbox{$N=8$} (upper panels) and \mbox{$N=12$} (lower panels) clusters, obtained from the ED calculations. 
  Inner white hexagons indicate the first Brillouin zone.%41 -> 49
  }
  \end{figure}

  Figure~\ref{fig:str-fac} presents 
  the momentum dependence of $S(\mathbf{q})/N$ 
  for several $\xi$ values within this competing regime, for \mbox{$N=8$} (top)
  and \mbox{$N=12$} (bottom).
  At the FM Kitaev point \mbox{($\xi=0.75$)}, $S(\mathbf{q})/N$ exhibits no pronounced peaks
  for either cluster, consistent with the Kitaev QSL.
  At the AKLT point \mbox{($\xi=1.0$)}, peaks appear at the $\Gamma^\prime$ points, but
  their intensities diminish from \mbox{$N=8$} to \mbox{$12$}---a hallmark of
  short-range AFM correlations in the AKLT VBS state.
  A distinct feature emerges slightly away from the FM Kitaev point, around \mbox{$\xi \sim 0.8$},
  where $S(\mathbf{q})/N$ develops peaks at the X points,
  suggesting stripy-like correlations (see Supplemental Material~\cite{SM}).
  In contrast, for \mbox{$0.85 \lesssim \xi <1.00$},
  neither cluster exhibits sharp peaks, and the overall magnitude of $S(\mathbf{q})/N$
  systematically decreases from \mbox{$N=8$} to \mbox{$12$}.
  Taken together, these results imply the suppression of long-range magnetic order in this competing regime,
  potentially yielding quantum disordered states. 
  The classical and semi-classical calculations also reveal strong competition among multiple states,
  suggesting that quantum fluctuations are sufficiently strong to suppress the classical orders predicted in this regime.

  %599

%---------------------------- Discussion ----------------------------%
\textit{Discussion.---}
  We place our results in the broader context of related spin models.
  First, we compare them with the \mbox{$S=3/2$} Kitaev-Heisenberg model on the honeycomb lattice,
  where the AKLT Hamiltonian in Eq.~\eqref{eq:HAKLT} is replaced with the Heisenberg model, 
  \mbox{$\hat{H}_{\rm Heis} = \sum_{\langle i,j \rangle} J\hat{\bm{S}}_i \cdot \hat{\bm{S}}_j$}~\cite{PhysRevB.106.174416, PhysRevResearch.6.033168, liu2025quantumphasediagramextended}.
  Previous studies on this model revealed a similar ground-state phase diagram,
  including the FM, zigzag, N\'eel, and Kitaev QSL phases, except in the
  competing regime between the FM Kitaev and AKLT limits \mbox{($0.75 < \xi < 1.00$)}.
  In this regime, the Kitaev-Heisenberg model predicts a stripy order for \mbox{$0.75 \lesssim \xi \lesssim 0.9$} 
  and a N\'eel order for \mbox{$0.9 \lesssim \xi < 1.0$}.
  While the stripy-like correlations near the FM Kitaev limit observed in our study may
  represent remnants of this stripy order, the suppressed spin correlations across this regime are
  attributed to the biquadratic and bicubic terms in the AKLT Hamiltonian, 
  highlighting the crucial role of higher-order multiple-spin interactions in the competing regime.

  At the classical and semi-classical levels, these additional interactions stabilize noncoplanar orders, as shown in Fig.~\hyperlink{fig:schematic_pd}{\ref{fig:schematic_pd}(c)}.  
  Similar $3Q$ orders were reported in related models, 
  including the \mbox{$S=1/2$} Kitaev model with additional ring exchanges~\cite{PhysRevB.108.014437}, 
  the \mbox{$S=1$} Kitaev model with bilinear-biquadratic interactions~\cite{PhysRevB.107.L140403, PhysRevResearch.6.033077}, 
  and more generally in \mbox{$O(N)$} models~\cite{jin2025effectivefieldtheorytripleq}.
  Importantly, such noncoplanar states are often fragile against quantum fluctuations and melt into quantum disordered phases~\cite{PhysRevB.96.115115, 10.21468/SciPostPhys.13.3.050}.
  Our results also follow this trend.

  Next, we compare our results with those for the \mbox{$S=1/2$} Kitaev-Heisenberg model on the star lattice \cite{PhysRevLett.99.247203, PhysRevLett.115.087203, PhysRevB.109.094421}.
  In this comparison, similar to the composite \mbox{$S=1/2$} picture for the AKLT VBS shown in Fig.~\hyperlink{fig:schematic_pd}{\ref{fig:schematic_pd}(b)}, 
  each \mbox{$S=3/2$} spin on the honeycomb lattice is viewed as three interacting \mbox{$S=1/2$} spins 
  located on each triangle of the star lattice.
  The \mbox{$S=1/2$} star-lattice model exhibits strong competition among singlet VBS, stripy order, and a chiral spin liquid within the corresponding region \mbox{$0.75 < \xi < 1.00$} in our model.
  This analogy suggests a possible connection between the potential quantum disordered states in our system and these competing phases, including VBS and chiral spin liquid.
  %440

%---------------------------- Summary and Perspectives ----------------------------%
  \textit{Summary and perspectives.---}
  %Summary.-
  We have investigated the ground-state phase diagram of the \mbox{$S=3/2$} honeycomb model,
  which interpolates between Kitaev QSL and AKLT VBS,
  employing three complementary approaches that incorporate quantum fluctuations at different levels.
  We showed that noncoplanar orders, stabilized within the classical and semi-classical approximations,
  are progressively suppressed as quantum fluctuations are incorporated,
  suggesting the emergence of a quantum-entangled state with featureless spin structure factor 
  in the competing regime between the Kitaev QSL and the AKLT VBS.

  Further characterization of the quantum-entangled state and its excitations 
  remains an important challenge beyond the scope of the present study. 
  In particular, full quantum calculations for larger system sizes, 
  employing complementary approaches such as density matrix renormalization group and infinite projected entangled pair states, 
  will be crucial to clarify what kinds of quantum-disordered phases are realized 
  and to determine the extent to which they persist in the AKLT VBS and N\'eel states.
  Additionally, exploring other Kitaev limits, including the Abelian and non-Abelian topological QSLs, 
  and elucidating their connections to the AKLT state represent compelling directions for future research.

  Our results highlight the intricate interplay between frustration and quantum fluctuations 
  in \mbox{$S=3/2$} honeycomb systems, 
  opening new avenues for exploring unconventional quantum phenomena.
  Such \mbox{$S=3/2$} systems are prevalent in diverse magnetic materials, 
  including transition metal compounds with $t_{2g}$ orbital degrees of freedom. 
  Notably, the \mbox{$S=3/2$} Kitaev interaction has been discussed in the context of $t_{2g}$ magnets 
  with strong spin-orbit coupling~\cite{Xu2018, Kim2019, PhysRevLett.124.017201, PhysRevResearch.3.013216}, 
  and the \mbox{$S=3/2$} AKLT VBS has also been proposed~\cite{PhysRevLett.114.247204}. 
  These observations suggest the intriguing possibility of realizing Kitaev-AKLT competition in real materials. 

  Last but not least, the emergence of a quantum-entangled disordered regime is 
  likely not specific to the Kitaev-AKLT interpolation. 
  More generally, competition between fractionalized spin-liquid correlations 
  and valence-bond ordering tendencies can enhance quantum fluctuations and 
  entanglement in intermediate parameter regions. The detailed nature and extent of this regime, 
  however, may depend sensitively on lattice geometry, symmetry, and the specific interactions involved.
  %258

\textit{Acknowledgments.---}
The authors thank Y. Kato for fruitful discussions.
S.I. was supported by the
Program for Leading Graduate Schools (MERIT-WINGS) and
Hirose Foundation.  
This work was supported by the JSPS KAKENHI (Grants
No. JP24K17009, No. JP25H01247, and No. JP25K17335).
Part of computation in this
work has been done using the facilities of the Supercomputer
Center, the Institute for Solid State Physics, the University of
Tokyo.

% The \nocite command causes all entries in a bibliography to bve printed out
%\nocite{*}

\bibliography{reference}% Produces the bibliography via BibTeX.

\begin{thebibliography}{68}
\bibitem [{ED_()}]{ED_note}%
  \BibitemOpen
  \href@noop {} {}\bibinfo {note} {For the $N=12$ cluster, several additional spikes appear near $\xi\sim 0.8$ in the second derivative of the ground-state energy. Since this cluster possesses lower momentum-space symmetry than the $N=8$ cluster, these features are most naturally attributed to finite-size level crossings associated with the reduced symmetry, rather than to additional phase boundaries.}\BibitemShut {Stop}%
\end{thebibliography}

%%%%%%%%%% Merge with supplemental materials %%%%%%%%%%
\pagebreak
%\appendix
\setcounter{page}{1}
\setcounter{equation}{0}
\setcounter{figure}{0}
\setcounter{table}{0}
\renewcommand{\theequation}{S\arabic{equation}}
\renewcommand{\thefigure}{S\arabic{figure}}
\setcounter{section}{0}
\setcounter{secnumdepth}{1}
\setcounter{secnumdepth}{2}

\begin{widetext}
\begin{center}
\textbf{\large Supplemental Material for \\
Kitaev Meets AKLT: Competing Quantum Disorder in Spin-3/2 Honeycomb Systems}

\vskip\baselineskip
Sogen Ikegami$^1$, Kiyu Fukui$^2$, Rico Pohle$^3$, and Yukitoshi Motome$^1$
\par
{$^1$\it Department of Applied Physics, the University of Tokyo, Tokyo 113-8656, Japan} \\
{$^2$\it Department of Physical Sciences, Ritsumeikan University, Kusatsu, Shiga 525-8577, Japan} \\
{$^3$\it Faculty of Science, Shizuoka University, Shizuoka 422-8529, Japan}
\end{center}
\end{widetext}

%---------------------------- Appendix A: Details of Methods ----------------------------%
\section{\label{sec:method}Details of methods}
In the main text, we employ three complementary methods. 
Here, we provide details of two of them: 
classical and semi-classical approaches. 

\subsection{\label{sec:classical}Classical: O(3) vector}
In the classical calculations, each \mbox{$S=3/2$} spin 
is treated as a classical \mbox{O(3)} vector $\bm{S}_i$. 
The classical spin Hamiltonian is obtained by replacing the spin operators in Eq.~\eqref{eq:H} with classical vectors, 
where the biquadratic and bicubic terms in Eq.~\eqref{eq:HAKLT} correspond to 
the square and cube of the inner product of these vectors. 
In models with only bilinear terms, the choice of the vector length affects only the overall energy scale. 
However, in the presence of higher-order terms, such as the biquadratic and bicubic terms,
the vector length influences their relative contributions to the total energy.
To ensure an appropriate classical limit, we fix the vector length to $|\bm{S}_i|=3/2$, which yields a %match the 
classical energy for the FM state %with the corresponding %ED result, which fixes the vector length to $|\bm{S}_i|=3/2$.
consistent with that for the quantum system.

Each classical spin is parametrized as
\begin{equation}
  \bm{S}_i(\theta_i, \phi_i) = \frac{3}{2} \left( \sin\theta_i \cos\phi_i, \sin\theta_i \sin\phi_i, \cos\theta_i \right),
\end{equation}
where \mbox{$\theta_i \in [0, \pi]$} and \mbox{$\phi_i \in [0, 2\pi)$} denote the polar and azimuthal angles, respectively.
To obtain the spin configurations for the lowest-energy state, we perform variational energy minimization
by optimizing the directions of classical \mbox{O(3)} vectors at each site, \mbox{$\{\theta_i,\phi_i\}$},
using the gradient descent method described in Sec.~\ref{sec:JAX}.

\subsection{\label{sec:SU4}Semi-classical: SU($N$) coherent state}
SU($N$) coherent states provide a powerful framework for representing quantum spins,
as they capture onsite quantum fluctuations beyond the classical \mbox{O(3)} vector picture.
In this formalism, the many-body wavefunction is expressed as a direct product
of local \mbox{SU($N$)} coherent states,
\begin{equation}
  \ket{\Psi_{\rm{coherent}}} = \bigotimes_i \ket{\Omega_i}_i,
\end{equation}
where the local \mbox{SU($N$)} coherent state \mbox{$\ket{\Omega_i}_i$} takes the form
\begin{equation}
  \ket{\Omega_i}_i = \sum_{\alpha=1}^{N} c_\alpha(\Omega_i) \ket{\alpha}_i,
\end{equation}
with the normalization condition \mbox{$\sum_{\alpha=1}^{N} |c_\alpha(\Omega_i)|^2 = 1$}.
Here, \mbox{$\{\ket{\alpha}_i\}$} denotes the basis of the local Hilbert space at site $i$,
and the coefficients \mbox{$\{c_\alpha(\Omega_i)\}$} parameterize the local quantum state.
For the \mbox{$S=3/2$} spin system, the local Hilbert space is four-dimensional ($N=4$),
spanned by the four $\hat{S}^z$ eigenstates: \mbox{$\{\ket{\alpha}\} = \{\ket{3/2}, \ket{1/2}, \ket{-1/2}, \ket{-3/2}\}$}.

The parametrization of the complex coefficients \mbox{$\{c_\alpha(\Omega)\}$} is not unique.
Following Refs.~\cite{Nemoto_2000, PhysRevB.108.L241108, pohle2025electronphonon},
we introduce a set of angular parameters,
\begin{equation}
  \Omega = (\theta_1, \cdots, \theta_{N-1}, \phi_1, \cdots, \phi_{N-1}),
\end{equation}
and define
\begin{equation}\label{eq:SU4_coeff}
  c_\alpha(\Omega) = e^{i\phi_{\alpha-1}} \cos \theta_\alpha \prod_{\beta=1}^{\alpha-1} \sin \theta_\beta,
\end{equation}
where \mbox{$\theta_\alpha \in [0, \frac{\pi}{2}]$} and \mbox{$\phi_\alpha \in [0, 2\pi)$}, with
the conventions \mbox{$\phi_0=\theta_N=0$} in Eq.~\eqref{eq:SU4_coeff}.
We optimize $\Omega$ to achieve the lowest-energy state using the method described in Sec.~\ref{sec:JAX}.

\subsection{\label{sec:JAX}Energy minimization by a gradient descent method}

To determine the ground-state phase diagrams shown in Fig.~\hyperlink{fig:schematic_pd}{\ref{fig:schematic_pd}(c)}, 
we perform large-scale variational energy minimization
using the machine learning library JAX~\cite{jax2018github}. 
In the classical \mbox{O(3)} vector approach,
the variational parameters consist of \mbox{$2N_{\rm site}$} angular variables 
\mbox{$\{\theta, \phi\}_i$} for \mbox{$N_{\rm site}$} spins on the honeycomb lattice.
In the semi-classical \mbox{SU(4)} coherent state approach,
each spin is parameterized by \mbox{$\Omega_i = \{\theta_1, \theta_2, \theta_3, \phi_1, \phi_2, \phi_3\}_i$},
corresponding to \mbox{$6N_{\rm site}$} variational parameters.

The minimization proceeds as follows.
We first generate an initial set of variational parameters and compute the energy $E$
together with its gradient with respect to these parameters.
The parameters are then updated along the gradient direction to reduce the energy.
By iterating this process until convergence of the energy and its gradient,
we obtain an energetically stable variational configuration.
The gradients are computed using automatic differentiation provided by JAX. 
For optimization, we employ the Optax library \cite{deepmind2020jax} with the Adam optimizer \cite{kingma2017adam}, adopting a learning rate of $0.1$.

To avoid bias in the optimization, it is essential to sample multiple initial configurations of 
variational parameters uniformly over the corresponding parameter space. 
For classical calculation, we generate \mbox{$\phi_i$} uniformly in \mbox{$[0, 2\pi)$},
and \mbox{$\theta_i$} from the uniform distribution as \mbox{$\theta_i = \arccos r_i$},
where \mbox{$r_i$} is a random number uniformly distributed in \mbox{$[0, 1]$}.
The semi-classical calculation requires a more sophisticated sampling scheme due to the higher-dimensional parameter space.
Following the uniform sampling scheme proposed in Ref.~\cite{pohle2025electronphonon},
we first generate random numbers \mbox{$r_\alpha$} uniformly in
\begin{equation}
  r_\alpha \in \left[0, \frac{1}{2(N-\alpha)} \right], \quad (\alpha=1,\ldots,N-1),
\end{equation}
and then convert them into \mbox{$\theta_\alpha$} as 
\begin{equation}
  \theta_\alpha = \arcsin \left\{ [2(N-\alpha)r_\alpha]^{1/(2(N-\alpha))} \right\},
\end{equation}
with uniformly generated \mbox{$\phi_\alpha$} in \mbox{$[0, 2\pi)$}.
The system sizes used in the calculations are \mbox{$N_{\rm site}= 2 \times 128 \times 128=32768$} for the classical approach
and \mbox{$N_{\rm site} = 2 \times 96 \times 96=18432$} for the semi-classical approach.

%---------------------------- Appendix: Noncoplanar order ----------------------------%
\section{\label{sec:3Q}Spin structures of noncoplanar orders}

\begin{figure}[t]
\hypertarget{fig:App_fig1}{}
\includegraphics[width=\columnwidth]{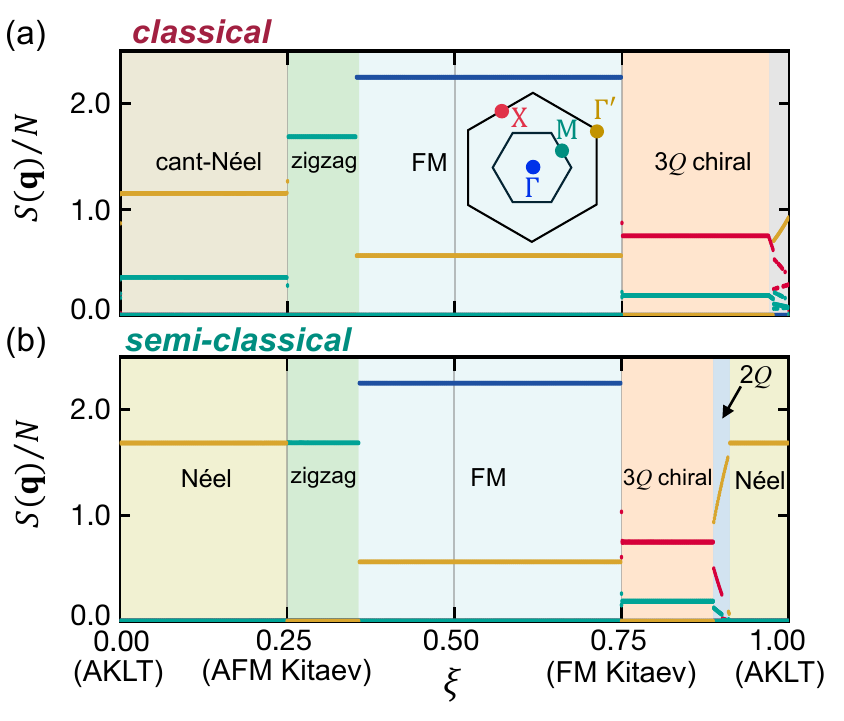}
\caption{\label{fig:App_fig1} 
$\xi$ dependence of the normalized spin structure factor, $S(\mathbf{q})/N$, at the $\Gamma$, M, X, and $\Gamma^\prime$ points, 
obtained by (a) classical and (b) semi-classical approaches.
The inset of (a) represents the Brillouin zone of the honeycomb lattice,
with the inner and outer hexagons indicating the first Brillouin zone and the extended zone up to the third one, respectively.
}
\end{figure}

\begin{figure}[t]
\hypertarget{fig:App_fig2}{}
\includegraphics[width=\columnwidth]{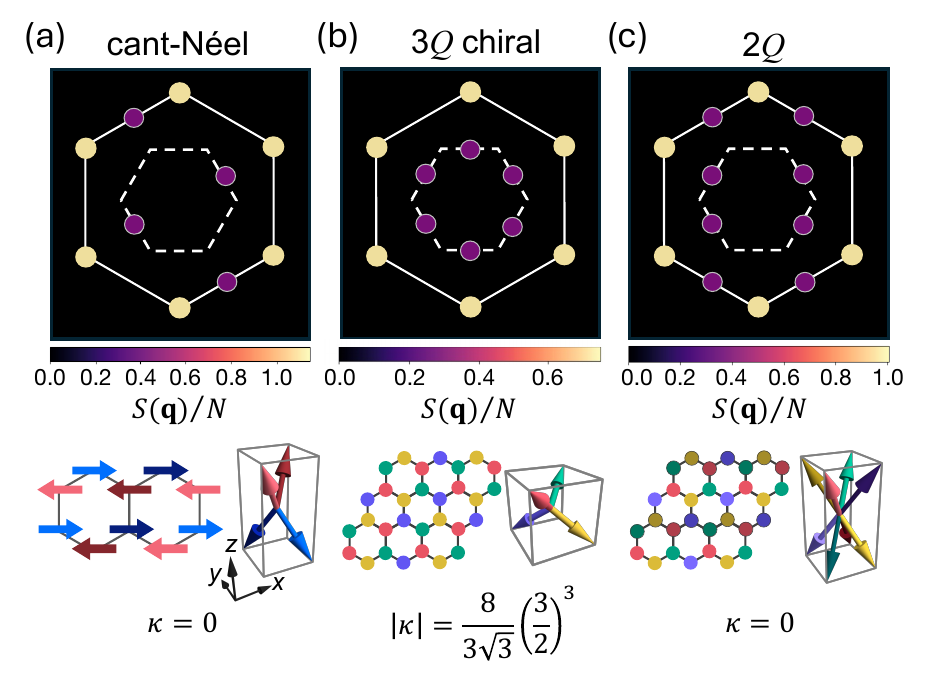}
\caption{\label{fig:App_fig2} 
Comparison of (a) cant-N\'eel, (b) 3$Q$ chiral, and (c) 2$Q$ orders. 
The upper panels show the momentum space distributions of \mbox{$S(\mathbf{q})/N$},
where the Bragg peaks are represented by circles for better visibility.
The lower panels display the corresponding real-space spin configurations 
along with the values of the scalar spin chirality \mbox{$\kappa$}.
}
\end{figure}

%In this section, we present the details of the 
In the main text, we identify three distinct noncoplanar orders
(cant-N\'eel, 3$Q$ chiral, and 2$Q$) in the classical and semi-classical results.
Here, we present their detailed spin structures.
Figure~\hyperlink{fig:App_fig1}{\ref{fig:App_fig1}} shows the corresponding plots to Fig.~\hyperlink{fig:EE}{\ref{fig:EE}(a)}, 
obtained from the classical and semi-classical calculations.
Figure~\ref{fig:App_fig2} displays the momentum space distributions of \mbox{$S(\bm q)/N$} (upper panels)
and the corresponding real-space spin configurations with their scalar spin chirality for each phase (lower panels).
The definition of the scalar spin chirality on the honeycomb lattice follows Ref.~\cite{PhysRevResearch.6.033077}.

While the semi-classical calculation stabilizes the N\'eel order for $0<\xi<0.25$,
the classical calculation instead yields the cant-N\'eel order, 
characterized by sharp peaks of \mbox{$S(\mathbf{q})/N$} at both the M and $\Gamma^\prime$ points,
as shown in Fig.~\hyperlink{fig:App_fig1}{\ref{fig:App_fig1}} and the upper-left panel of Fig.~\hyperlink{fig:App_fig2}{\ref{fig:App_fig2}}.
This phase features an enlarged magnetic unit cell with four sublattices,
where the spins point toward the vertices of an elongated tetrahedron, 
as shown in the lower-left panel of Fig.~\hyperlink{fig:App_fig2}{\ref{fig:App_fig2}}.
The real-space spin configuration of this phase is viewed as a noncoplanar canting deformation
of the collinear N\'eel state, while maintaining zero scalar spin chirality.
The classical framework evaluates energy through inner vector products, 
with biquadratic interactions penalizing parallel spins, 
thereby favoring noncoplanar cant-Néel states. 
In contrast, the semi-classical approach accounts for onsite quantum fluctuations, 
leading to markedly different contributions from higher-order interactions. 
Quantum corrections lower the energy of collinear configurations and suppress noncoplanar ordering, 
ultimately stabilizing collinear Néel order—opposite to the classical prediction.

Both classical and semi-classical calculations identify the 3$Q$ chiral order for $0.75<\xi \lesssim 0.98$ and $0.75<\xi \lesssim 0.88$, respectively.
This state exhibits the strongest intensities of \mbox{$S(\mathbf{q})/N$} at all $\Gamma^\prime$ points
and the second strongest ones at all M points [Fig.~\hyperlink{fig:App_fig1}{\ref{fig:App_fig1}} 
and upper panel of Fig.~\hyperlink{fig:App_fig2}{\ref{fig:App_fig2}(b)}].
In real space, the 3$Q$ chiral order forms a four-sublattice magnetic unit cell,
where the spins point toward the vertices of a regular tetrahedron, 
giving rise to a nonzero scalar spin chirality [lower panel of Fig.~\hyperlink{fig:App_fig2}{\ref{fig:App_fig2}(b)}] \cite{PhysRevLett.79.2081, doi:10.1143/JPSJ.79.083711, PhysRevB.83.184401, PhysRevResearch.6.033077}.

In the region \mbox{$0.88 \lesssim \xi \lesssim 0.90$}, the semi-classical calculation reveals the 2$Q$ order.
This state shows strong peaks at all $\Gamma^\prime$ points
and weaker peaks at four out of the six M points in \mbox{$S(\mathbf{q})/N$} 
(Fig.~\hyperlink{fig:App_fig1}{\ref{fig:App_fig1}} and upper panel of Fig.~\hyperlink{fig:App_fig2}{\ref{fig:App_fig2}(c)}).
In real space, the 2$Q$ order features an enlarged magnetic unit cell with eight sublattices,
where the spins point toward the vertices of a cuboid (lower panel of Fig.~\hyperlink{fig:App_fig2}{\ref{fig:App_fig2}(c)}).
Each plaquette carries a nonzero scalar spin chirality; however, the signs alternate from plaquette to plaquette,
resulting in a vanishing total chirality.
As shown in Fig.~\hyperlink{fig:App_fig1}{\ref{fig:App_fig1}},
the relative intensities of \mbox{$S(\mathbf{q})/N$} and the angles among the eight-sublattice spins
vary continuously within this 2$Q$ phase,
while they remain independent of \mbox{$\xi$} for both the cant-N\'eel and 3$Q$ chiral orders.
In contrast, the classical framework fails to yield
converged solutions in the region between the 3$Q$ chiral and canted N\'eel phases for \mbox{$0.98\lesssim \xi < 1.00$}.

%---------------------------- Appendix: GSWT ----------------------------%
\section{\label{sec:GSWT}Generalized spin-wave theory}

\begin{figure}[t]
\hypertarget{fig:App_fig3}{}
\includegraphics[width=\columnwidth]{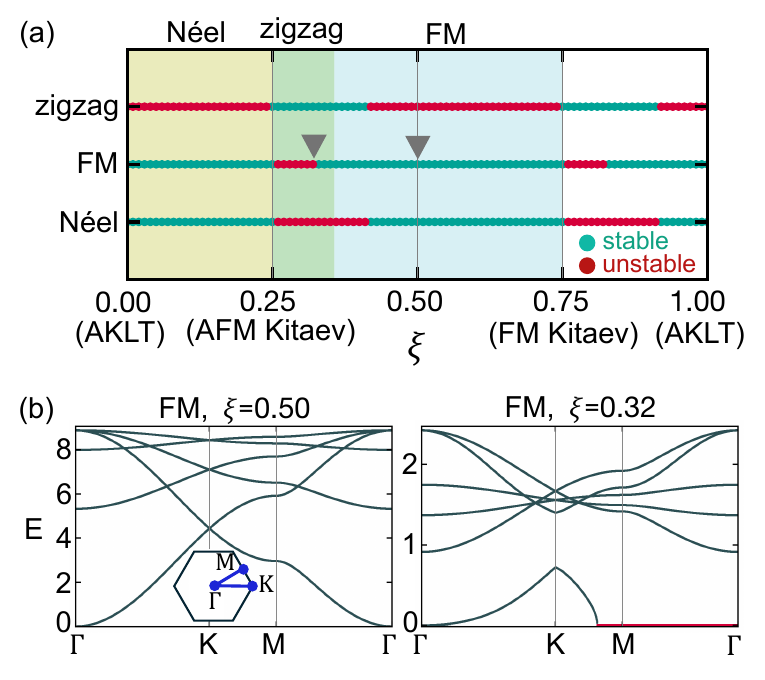}
\caption{\label{fig:App_fig3} 
(a) Stability of the FM, N\'eel, and zigzag ordered phases examined using GSWT.
The yellow, green, and blue shaded regions indicate the 
N\'eel, zigzag, and FM phases, respectively, as determined from
the semi-classical ground-state phase diagram in Figs.~\protect\hyperlink{fig:schematic_pd}{\ref{fig:schematic_pd}(c)} and \protect\hyperlink{fig:App_fig1}{\ref{fig:App_fig1}(b)}.
The green and red markers indicate whether each state is stable or unstable within GSWT.
The two gray triangles mark the parameter values used in (b).
(b) Spin-wave excitation spectra calculated by GSWT at \mbox{$\xi=0.5$} (left panel) and \mbox{$0.32$} (right panel) from the FM ordered state
along the high symmetric lines of the Brillouin zone, as illustrated in the inset.
The red line in the right panel represents nonzero imaginary components in the excitation spectra, 
signaling the instability of the FM order at this parameter.
}
\end{figure}

We here analyze the stability of FM, N\'eel, and zigzag ordered phases, identified in the semi-classical %results
phase diagram, using the generalized spin-wave theory (GSWT) \cite{10.1093/ptep/ptu109}.
Unlike the conventional spin-wave theory based on the Holstein-Primakoff transformation, which examines %analyzes
stability by expanding local \mbox{SU(2)} fluctuations around a given state,
the GSWT extends this approach by incorporating \mbox{SU($N$)} fluctuations.
The key step involves introducing $N$-flavor Schwinger bosons at each site,
and expressing the \mbox{SU($N$)} generators as bilinear forms of these bosons.
This allows GSWT to capture onsite quantum fluctuations beyond the \mbox{SU(2)}-based spin-wave theory,
which is crucial for higher-spin systems with higher-order interactions, such as the biquadratic and bicubic terms in the current model.

Figure \hyperlink{fig:App_fig3}{\ref{fig:App_fig3}(a)} displays the stability of 
the zigzag, FM, and N\'eel ordered phases, shown
from the top to bottom rows, respectively.
The green and red markers indicate whether each ordered state is stable or unstable within GSWT.
The yellow, green, and blue shaded regions correspond to the N\'eel, zigzag, and FM orders, respectively,
in the semi-classical phase diagram of Figs.~\hyperlink{fig:schematic_pd}{\ref{fig:schematic_pd}(c)} and \hyperlink{fig:App_fig1}{\ref{fig:App_fig1}(b)}.
Overall, the GSWT reproduces the phase stability well, except 
near the boundary between the zigzag and FM phases, 
where it overestimates the stability of both orders. 

Figure \hyperlink{fig:App_fig3}{\ref{fig:App_fig3}(b)} exemplifies the spin-wave excitation spectra
calculated by GSWT at \mbox{$\xi=0.5$} (left panel) and \mbox{$0.32$} (right panel) from the FM ordered state
along the high symmetric lines of the Brillouin zone, as indicated in the inset.
At \mbox{$\xi=0.5$}, the excitation spectra show quadratic dispersions at the $\Gamma$ point, 
characteristic of the FM ordered state.
In contrast, for \mbox{$\xi=0.32$}, the spectra include
nonzero imaginary components along
the red line, signaling the instability of the FM order at this parameter.

%---------------------------- Appendix: Anistropic perturbation ----------------------------%
\section{\label{sec:anist}Effect of anisotropy}

\begin{figure}[thp]
\hypertarget{fig:App_fig4}{}
\includegraphics[width=\columnwidth]{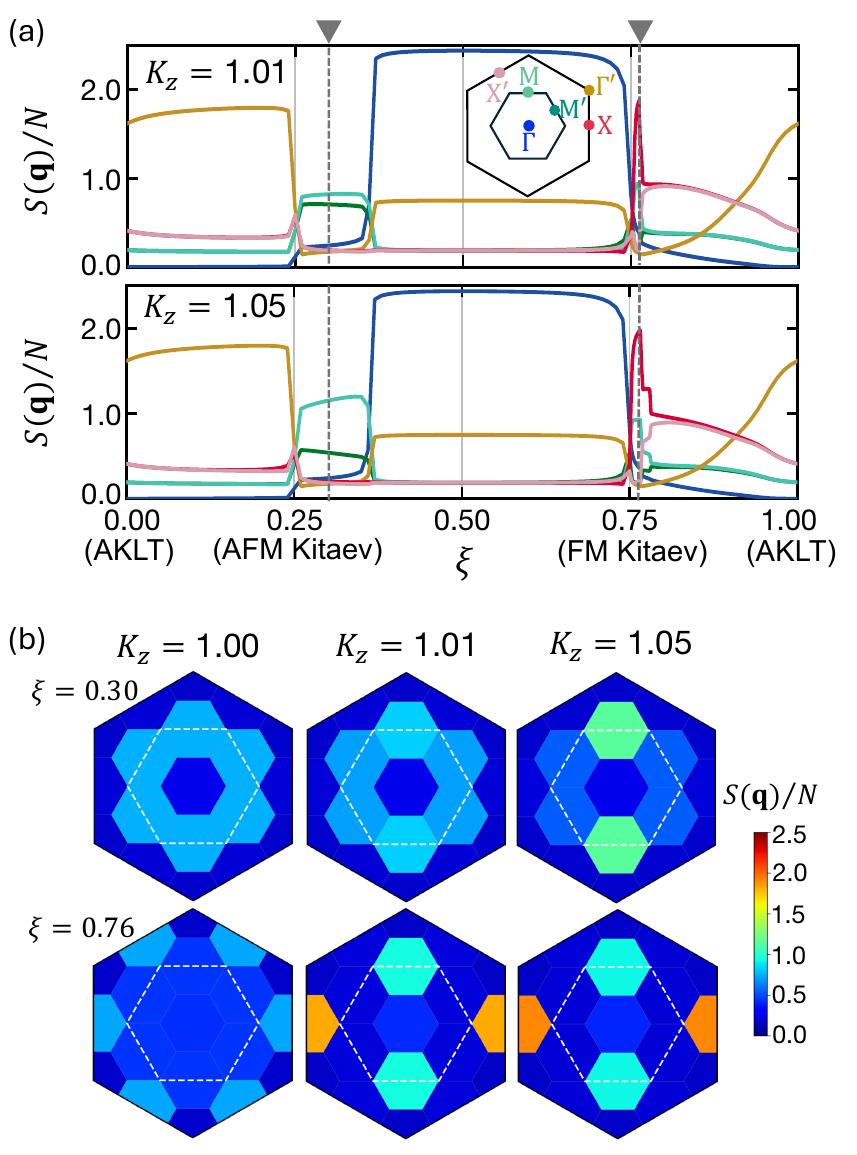}
\caption{\label{fig:App_fig4} 
(a) $\xi$ dependence of the normalized spin structure factor, \mbox{$S(\mathbf{q})/N$}, at high-symmetry points.
The upper and lower panels show the results for \mbox{$K_z=1.01$} and \mbox{$1.05$}, respectively, with \mbox{$K_x=K_y=J=1$}.
The two gray dashed lines indicate the parameter values used in (b).
(b) Momentum space distributions of \mbox{$S(\mathbf{q})/N$}
up to the third Brillouin zone at \mbox{$\xi=0.3$} (upper panels) and \mbox{$\xi=0.76$} (lower panels)
for \mbox{$K_z=1.00$}, \mbox{$1.01$}, and \mbox{$1.05$}. 
The inner dashed white hexagons indicate the first Brillouin zone.
The results are obtained from ED for \mbox{$N=8$} cluster.
}
\end{figure}

For the model employed in the main text,
the spin structure factor obtained from ED calculations on the \mbox{$N=8$} cluster
exhibits identical intensities across all six M points and all six X points,
as shown in Figs.~\hyperlink{fig:EE}{\ref{fig:EE}(a)} and \ref{fig:str-fac}.
In small clusters with high symmetry, ED calculations often yield a ground state
that is a cat state,
i.e., a superposition of symmetry-related degenerate states.
This makes it challenging to infer the intrinsic correlations in the system.
To reveal these intrinsic tendencies, it is useful to introduce weak perturbations that explicitly break the symmetry.
Here, we present ED results for the \mbox{$N=8$} cluster under weak anisotropy in the Kitaev interactions.
Specifically, we implement the anisotropy by introducing a bond-dependent Kitaev coupling in Eq.~\eqref{eq:HKitaev} as
\begin{equation}\label{eq:HKitaev2}
  \hat{H}_{\rm{Kitaev}} = \sum_{\gamma= x,y,z} \sum_{\langle i,j \rangle_\gamma}K_\gamma \hat{S}_i^\gamma \hat{S}_j^\gamma.
\end{equation}
In the calculations below, we set \mbox{$K_z=1.01$} and \mbox{$K_z=1.05$}, while keeping \mbox{$K_x=K_y=J=1$}.

Figure~\hyperlink{fig:App_fig4}{\ref{fig:App_fig4}(a)} 
shows the $\xi$ dependency of \mbox{$S(\mathbf{q})/N$} at high symmetry momenta
for \mbox{$K_z=1.01$} (upper panel) and \mbox{$K_z=1.05$} (lower panel). 
Figure~\hyperlink{fig:App_fig4}{\ref{fig:App_fig4}(b)} displays
the momentum space distributions of \mbox{$S(\mathbf{q})/N$} 
at \mbox{$\xi=0.3$} (upper panels) and \mbox{$\xi=0.76$} (lower panels) for \mbox{$K_z=1.00$}, \mbox{$1.01$}, and \mbox{$1.05$}. 
As shown in these figures, within the range \mbox{$0.25 \lesssim \xi \lesssim 0.36$}, even a small anisotropy in the Kitaev interactions
significantly enhances the intensities of the spin structure factor
at two of the six M points relative to the others.
This anisotropy becomes more pronounced as \mbox{$K_z$} increases,
indicating that the underlying correlations in the isotropic system
exhibit zigzag AFM character in this regime.
Similarly, for \mbox{$0.75 \lesssim \xi \lesssim 0.8$},
the intensities at two of the six X points and two of the six M points grow with increasing \mbox{$K_z$},
signaling intrinsic stripy correlations.
In contrast, the region \mbox{$0.8 \lesssim \xi \lesssim 1.0$} shows little change in \mbox{$S(\mathbf{q})/N$}
with \mbox{$K_z$}, suggesting that the nature of
this possible quantum disordered state differs
qualitatively from that in \mbox{$0.75 \lesssim \xi \lesssim 0.8$}. %, where stripy type spin correlations are immediately enhanced

\end{document}